# The ERIS Adaptive Optics System

A. Riccardi*[a,c], S. Esposito[a,c], G. Agapito[a,c], J. Antichi[a], V. Biliotti[a,c], C. Blain[a], R. Briguglio[a,c], L. Busoni[a,c], L. Carbonaro[a,c], G. Di Rico[a,c], C. Giordano[a,c], E. Pinna[a,c], A. Puglisi[a,c], P. Spanò[a], M. Xompero[a,c], A. Baruffolo[b], M. Kasper[d], S. Egner[d], M. Suàrez Valles[d], C. Soenke[d], M. Downing[d], J. Reyes[d]

[a]INAF-Osservatorio Astrofisico di Arcetri, Largo E. Fermi 5, 50125 Firenze, Italy
[b]INAF-Osservatorio Astronomico di Padova, Vicolo dell'Osservatorio, 5, 35141 Padova PD, Italy
[c]ADONI – Laboratorio Nazionale di Ottica Adattiva, Italy
[d]European Southern Observatory, Karl-Schwarzschild-Strasse 2, D-85748 Garching, Germany

**ABSTRACT**

ERIS is the new AO instrument for VLT-UT4 led by a Consortium of Max-Planck Institut fuer Extraterrestrische Physik, UK-ATC, ETH-Zurich, ESO and INAF. The ERIS AO system provides NGS mode to deliver high contrast correction and LGS mode to extend high Strehl performance to large sky coverage. The AO module includes NGS and LGS wavefront sensors and, with VLT-AOF Deformable Secondary Mirror and Laser Facility, will provide AO correction to the high resolution imager NIX (1-5um) and the IFU spectrograph SPIFFIER (1-2.5um). In this paper we present the preliminary design of the ERIS AO system and the estimated correction performance.

## 1. INTRODUCTION

ERIS, the Enhanced Resolution Imager and Spectrograph, is a new 1-5 μm instrument for the Cassegrain focus of the UT4/VLT telescope that is equipped with the Adaptive Optics Facility (AOF)[1]. The instrument is led by a Consortium of Max-Planck Institut für Extraterrestrische Physik (MPE, leading institute), UK Astronomy Technology Centre (UK-ATC), Swiss Federal Institute of Technology (ETH-Zurich), European Southern Observatory (ESO) and Istituto Nazionale di Astrofisica (INAF-Arcetri, Teramo and Padova). The ERIS Consortium have taken the lead of the project in early 2015 and successfully passed the Preliminary Design Review (PDR) in February 2016. The project is currently in Final Design phase and the related review is foreseen in March 2017. The present paper reports the design of the ERIS AO system as result of the PDR.

The instrument is made of the following sub-systems (see Figure 1):

- two science instruments, which receive their light via a IR/VIS dichroic beam-splitter located in the AO module.
  - NIX[4] provides diffraction limited imaging, sparse aperture masking (SAM) and pupil plane coronagraphy capabilities from 1-5 μm (i.e. J-M'), either in "standard" observing mode or with "pupil tracking" and "burst" (or "cube") readout mode. NIX is a cryogenic instrument and it is equipped with a 2048 × 2048 detector providing a field of view of 53" x 53"
  - SPIFFIER[5] is an upgraded version of SPIFFI, the 1-2.5 μm integral field unit used on-board SINFONI, that will be modified to be integrated into ERIS. Its observing modes are identical to those of SINFONI, possibly adding a high-resolution mode as a goal.
  - The two science instruments do not operate simultaneously, an insertable mirror behind the IR/VIS dichroic allows to select the instrument.
- the AO module has wavefront sensing and real-time computing capabilities. It interfaces to the AOF infrastructure and it is required to provide the following observing modes:

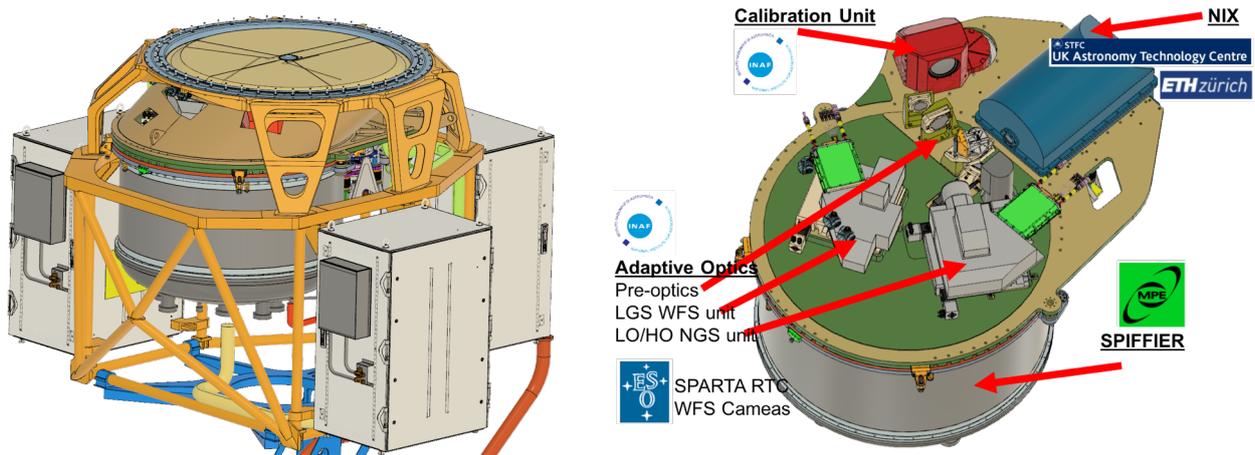

*Figure 1 Left: the ERIS instrument. Right: the ERIS sub-systems (optical bench internal view).*

- o LGS-mode: a Wavefront Sensor (WFS) provides high-order AO correction using a LGS on-axis and a second WFS provides low-order correction using a NGS in the patrol field (R≤1arcmin).
- o NGS-mode: a WFS provides high-order AO correction using a NGS in the patrol field;
- o Seeing Enhanced mode: only the on-axis LGS WFS is used for the high-order correction. This mode is used when no suitable NGS is available for the fast correction of the tip-tilt.
- The Calibration Unit (CU)[6], which provides facilities to calibrate the scientific instruments (remove instrument signature) and perform troubleshooting and periodic maintenance tests of the AO modules.

The ERIS AO concept maximizes the re-use of existing AOF sub-systems and components as explicitly requested by the ERIS Top Level Requirements (TLRs). In particular, the AO correction is provided by the AOF Deformable Secondary Mirror (DSM)[1][2] and the artificial sodium Laser Guide Star (LGS) is generated by one of the launchers of the 4LGSF system[3]. Moreover the ERIS AO sub-system uses components from the AOF GLAO systems (GALACSI and GRAAL), namely the wavefront sensor camera detectors (EMCCD220) and a modified version of SPARTA/AOF as Real-Time Computer (RTC).

## 2. LAYOUT

Figure 2 shows the conceptual scheme of the ERIS instrument, including the Adaptive Optics (AO) sub-system (or module). The set of optical components, that relays the telescope optical beam out from the Cassegrain Flange to the science instruments and to the AO WFSs, is defined as Warm Optics (WO) and is considered part of the AO sub-system.

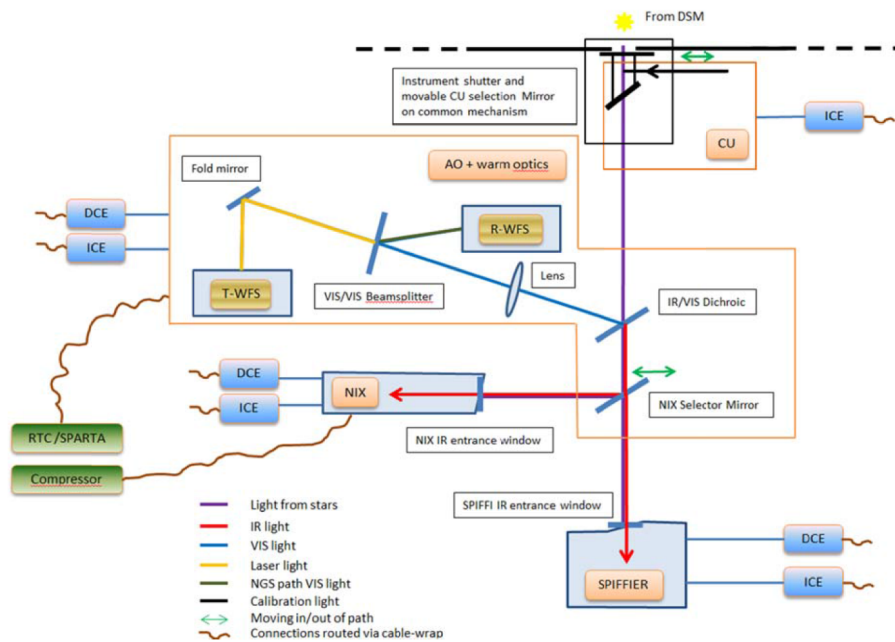

*Figure 2 ERIS instrument scheme*

The ERIS instrument is installed at the UT4/VLT Cassegrain Focal Station. The beam from the telescope is split by an IR/VIS dichroic, transmitting the IR (1-5μm) light to the science instruments and the visible (<1μm) to the AO WFSs. The IR/VIS dichroic is the first component of the WO and has also the practical function of decoupling the alignment of the AO beam from the science beam. The size of the dichroic (110mm) is constrained by the reflected AO patrol field of 2 arcmin diameter.

The two instruments (SPIFFIER and NIX) are simultaneously integrated in ERIS. SPIFFIER receives directly the beam from the IR/VIS dichroic. NIX is fed by inserting a 45deg mirror, the NIX Selector Mirror, still part of the WO.

The AO module does not need to internally integrate any deformable mirror, interfacing for the wavefront correction to the external DSM facility, part of the AOF[1][2].

The LGS-mode requires the implementation of two WFSs, the HO LGS WFS and the LO NGS WFS, requiring the introduction in the WO of a VIS/VIS Beam-Splitter in the optical branch reflected by the IR/VIS Dichroic. It separates the narrow-band Sodium LGS light from the rest of the visible radiation. Hereafter, we will refer to the VIS/VIS beam-splitter also as WFS Dichroic. The current design joins the HO NGS functionality in the LO NGS WFS, separating the functionalities between the two WFSs in a purely on-axis LGS-dedicated WFS and an HO/LO NGS-dedicated WFS with off-axis capabilities. The LGS WFS is implemented in the beam transmitted through the WFS dichroic because the HO LGS loop is less sensitive with respect to the HO NGS loop to the astigmatism and chromaticity introduced by the transmission through the WFS dichroic substrate.

The AO WFSs opto-mechanical design has been constrained by the general design choice of using flat (or extremely slow) optics to relay the telescope beam to the AO WFSs to maximize the stability of Non Common Path Aberrations (NCPA). This is an opportunity, to be exploited, provided by telescopes with an adaptive secondary mirror, forcing no powered optics to integrate a DM in the AO module. As a direct consequence of this design choice, the selection of the Natural Guide Star (NGS) in the AO field is not performed by an optical field selector (tip-tilt mirror on a pupil image), but implementing each WFS units as a compact board supported by stages to scan the field (circular, about 32mm radius). The same approach is adopted for tracking the focus in both NGS and LGS WFS case. This design choice is the standard approach for the AO systems developed by INAF-OAA for telescopes with Adaptive Secondary Mirrors and successfully experienced on LBT-FLAO[7] and Magellan-AO (MagAO)[8].

The mechanical scanning of the WFS board to patrol the field requires inserting a telecentric lens between the IR/VIS and VIS/VIS dichroic to relay the telescope exit pupil (M2) to infinity and keep the chief ray of the beam parallel regardless the off-axis. This is an extremely slow optics having the focal length equal to the distance to M2 (14.9 m).

The CU Selector Mirror is located upstream the IR/VIS Dichroic to allow the CU providing all the calibration needs to all the sub-systems. The size of the CUSM is driven by the NIX FoV and contributes to constrain the available space for the WO components. The distance between the Cassegrain Flange and the top of SPIFFIER provides another major space constraint to fit the WO design. That requires to increase the Cassegrain Back Focal Length (BFL) by 250 mm with respect to the nominal VLT one.

### 3. HO/LO NGS WFS UNIT

Figure 3 shows the design of the baseline NGS WFS implementation. It merges the HO and LO NGS WFS functionalities for the NGS and LGS mode, respectively. The HO configuration implements a 40x40 SH, i.e. the highest order compliant to the SPARTA/AOF and CCD220 format constraints. The LO configuration is implemented as a 4x4 SH array as a trade-off between pushing the performance toward the faint-end and allowing the sensing of a suitable number low orders modes for the truth sensing. Figure 4 shows the variation of the modal

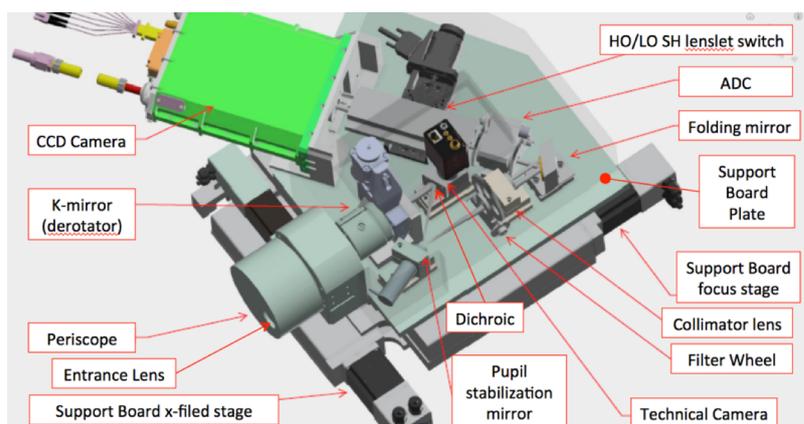

*Figure 3. NGS WFS unit. The unit cover is hidden to show the content.*

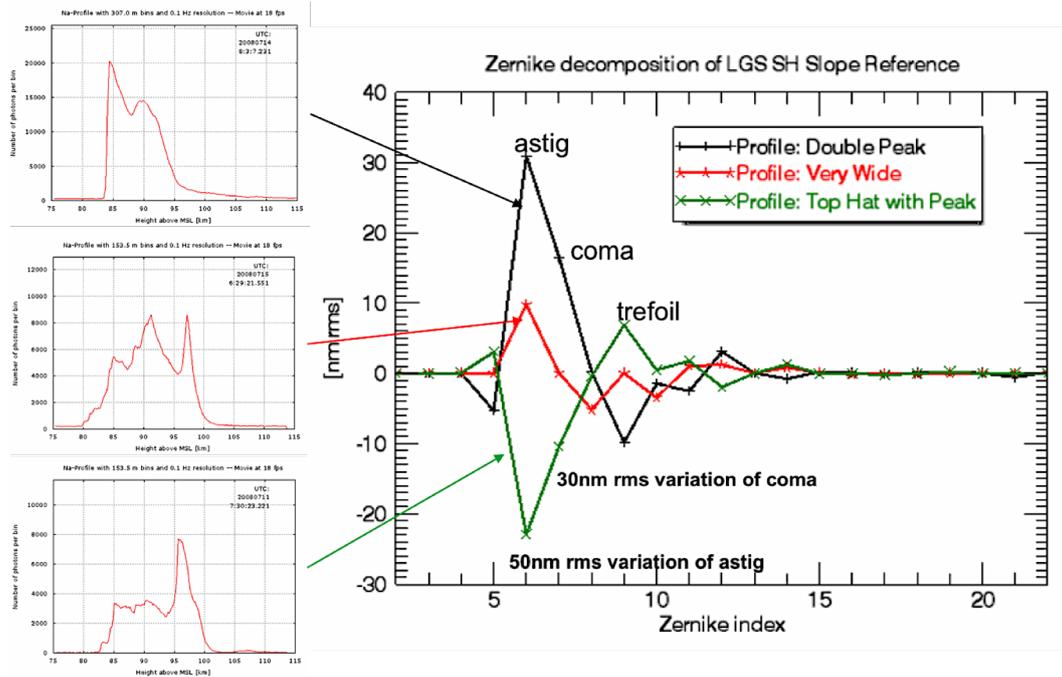

*Figure 4 Modal reconstruction error using a reference reconstructor with different sodium profiles. No atmospheric turbulence is present in the simulation. The beacon is re-centered and the LGS WFS is re-focused to simulate the correction effect of the LGS jitter and focus loops. The error has sensitive components up to the trefoil Zernike mode. Sodium profiles from [9]*

reconstruction error of an unaberrated wavefront from the HO LGS sensor signal when the sodium density profile changes its vertical distribution. The error is dominated by the Zernike mode up to the trefoil, requiring 4x4 SH as a suitable sampling for the truth sensing. Numerical simulations show that the use of a 4x4 configuration with respect to the minimal 2x2 produces only a 0.5 mag loss of the limit magnitude in the LGS mode.

The main components and functions of the LO/HO WFS unit are:

1. A dual axis stage supporting the WFS board. One (focus) stage moves in the direction of the input optical axis, the other (X stage) in the orthogonal direction, parallel to the plane of the WFS board. The focus stage is used to compensate for differential focus with respect to SPIFFIER and NIX, including drifts due to differential flexures and thermal expansion. The X stage, together with the periscope (see below), is used to patrol the NGS FoV (R=1' or R=32mm) and to compensate for all the effects introducing a differential image drift at the WFS focal plane with respect to the instrument focal plane. Part of the travel budget of the stages is used for the initial optical alignment of WO and WFS units to the ERIS reference axis.
2. A board, supported by the dual axis stage, where all the WFS components are mounted. The board can be separated from the stage for maintenance; the alignment is reproduced by a set of mechanical references.
3. A rotary stage supporting the Entrance Lens and a periscope. The axis offset of the periscope is 32 mm, allowing to patrol the full NGS FoV combining the periscope rotation with the X stage motion. The periscope solution has been triggered by the vertical space constraints inside the ERIS central structure that does not allow implementing the vertical (Y) motion with a third stage.
4. The Entrance Lens produces a F/20 beam and relays the input telecentric focal plane to a tip-tilt mirror for pupil stabilization purposes (see below).
5. A Pupil Rotator (named also K-mirror), implemented with an Abbe rotator prism, to de-rotate the actuator pattern of the DSM on the SH lenslet array when ERIS instrument rotator is operating in field tracking mode.
6. A beam splitter dichroic transmitting wavelengths shorter than 600nm to a technical camera that can be also used for NGS acquisition. The wavelengths in the range 600nm-1000nm are reflected toward the pupil stabilization mirror.
7. The pupil stabilization mirror on the F/20 focal plane used for stabilize the lateral decentering of the pupil on the SH lenslet array. The main contributors to the stroke requirement of the mirror are the pupil plane lateral drift due to alignment error between WFS and instrument rotator axis, PSF wobbling during K-mirror rotation due to residual alignment errors, flexures and thermal effects, and possible telescope exit pupil shift. The mirror implements a mask for a 2.5" field stop.
8. A 4-stop filter wheel hosting a beam stop for dark calibration, a pass-all filter for normal operations and two filters to reduce the light level and avoid CCD220 over-illumination with bright NGSs (1<mR<2 and 1<mR<7 in the HO and LO configurations)
9. A collimator lens relaying an image of the pupil on the SH lenslet array.
10. An ADC to compensate for atmospheric dispersion down to z=70deg and avoiding sensitivity reduction due to PSF elongation in the LO configuration.

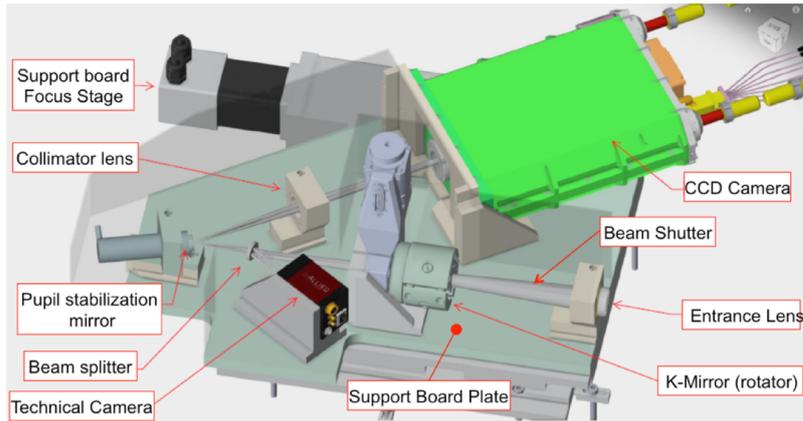

*Figure 5. NGS WFS unit. The unit cover is hidden to show the content.*

11. A stage to switch between HO and LO SH lenslet arrays and related optics relaying the SH spots to the WFS Camera CCD. The HO and LO channels are implemented as two parallel barrels with a 40x40 and 4x4 SH arrays.
12. The ESO's WFS Camera (CCD220) having 240x240 pixel format. It is the same camera used for AOF, providing 6x6 pixel per subaperture in the HO configuration (2.5"/6=0.42"/pix). In the LO configuration the 2.5" FoV sampling is currently 12x12 pixel per subaperture (0.21"/pix).

## 4. LGS WFS UNIT

Figure 5 shows the design of the baseline LGS WFS implementation. The main components and functions are:

1. A focus stage supporting the WFS board. The stage is used for tracking the best focus of laser beacon when changing its distance with elevation and sodium profile evolution. The travel of the stage is also used for compensating differential focus positions with respect to SPIFFIER and NIX, including the focus drifts due to differential flexures and thermal expansion.
2. A board, supported by the focus stage, where all the WFS components are mounted. The board can be separated from the stage for maintenance; the alignment is reproduced by a set of mechanical references.
3. An Entrance Lens producing a F/20 beam and relaying the input focal plane to a tip-tilt mirror for pupil stabilization purposes.
4. A shutter to stop the incoming beam to protect CCD220 against over-illumination and calibrate dark frames.
5. A Pupil Rotator (the same as NGS WFS) to de-rotate the actuator pattern of the DSM on the SH lenslet array when ERIS instrument rotator is operating in field tracking mode.
6. A beam splitter reflecting 1% of the laser light to feed a technical camera that can be also used for LGS acquisition. This camera will be probably removed because of the recently proven good performance of the 4LGSF pointing system[3].
7. The pupil stabilization mirror on the F/20 focal plane used for stabilize the lateral decentering of the pupil on the SH lenslet array. The main contributors to the stroke requirement of the mirror are the pupil plane lateral drift due to alignment error between WFS and instrument rotator axis, K-mirror and flexures. The mirror implements a mask for a 5.0" field stop to accommodate the elongated image of the sodium beacon.
8. A collimator lens relaying an image of the pupil on the SH lenslet array.
9. The ESO's WFS Camera (CCD220) including a 40x40 lenslet array glued on the camera window with the same size of the CCD (5.76 mm). It is the same camera and SH array used for AOF, providing 6x6 pixel per subaperture with a FoV of 5".

The LGS WFS operates only on-axis, therefore it does not require stages to patrol the field. LGS on-axis location and stabilization are devoted to the 4LGSF launcher, providing an internal large stroke field steering mirror for LGS acquisition and drift tracking, and a low stroke fast TT corrector.

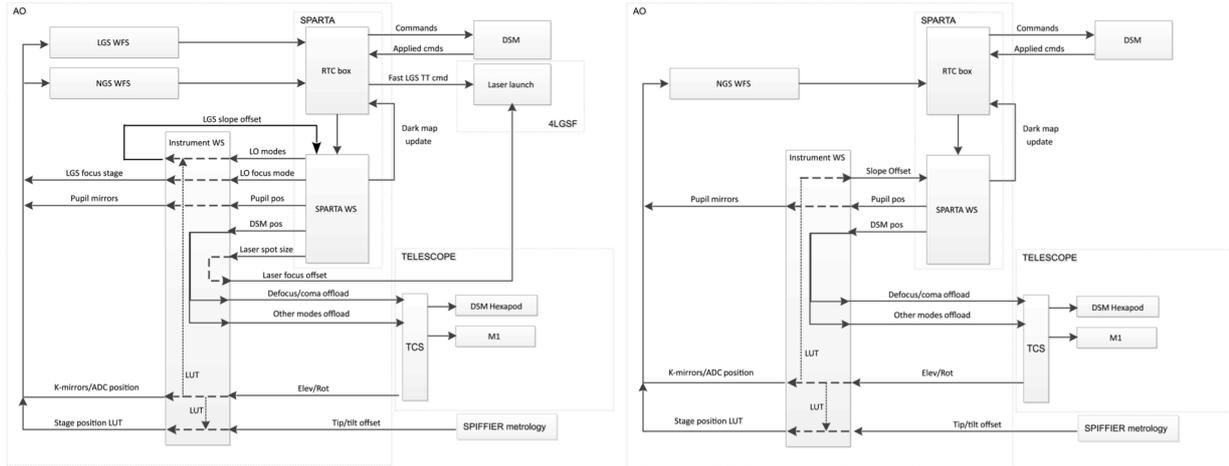

*Figure 6. Scheme of the real-time and auxiliary control loops in LGS mode(left) and NGS mode (right)*

## 5. REAL-TIME AND AUXILIARY CONTROL LOOPS

### 5.1 Control loops in LGS-mode

Figure 6-left summarizes the system of control loops and look-up tables (LUTs) in LGS mode. Both the LGS and NGS WFSs are operating in this mode. The NGS WFS is set in the LO configuration.

The real-time AO loop between the LGS WFS signal, the DSM and the laser launcher jitter mirror (part of the 4LGSF system) run with a baseline maximum loop frequency of 1 kHz.

The frames of the LGS WFS Camera are collected by SPARTA and processed to compute the DSM command vector. The DMS command vector does not include the tip-tilt component that is used to produce the laser jitter mirror command to stabilize the LGS image in the WFS. SPARTA sums the DSM command vector to the last available DSM tip-tilt vector from the LO loop and send it to the DSM. SPARTA receives back the last successfully applied command to manage the DSM saturation check.

The frames of the NGS WFS Camera in the LO 4x4 configuration are collected by SPARTA and processed to compute the DSM tip-tilt vector implementing the Active Vibration Cancellation algorithm developed by ESO[10]. The higher orders terms are also reconstructed and time averaged by SPARTA for the truth sensing loops. In particular the average NGS focus is used to relocate the LGS WFS focus stage and the higher orders to update the slope offset vector of the LGS WFS signals.

Among the auxiliary loops we cite here, for brevity, only a subset of the most relevant: a) the pupil position loop to keep pupil registered with the related SH arrays; b) the DSM averaged position loop to offload integrated focus/coma and higher orders to M2 hexapod and M1 active optics respectively; c) the k-mirror and ADC angles following the updating of the telescope elevation and instrument rotation angles; d) the image motion compensation loop updating an offset of the NGS WFS stages and periscope to compensate – with a pre-calibrated LUT as a function of elevation, instrument rotator angle and temperature – for the differential flexure between NGS WFS and the instrument, the NGS WFS k-mirror and periscope PSF induced wobble and the differential atmospheric refraction (NGS vs science target). When SPIFFIER is used, its internal metrology system can be used for computing an additional term in the differential flexure offset.

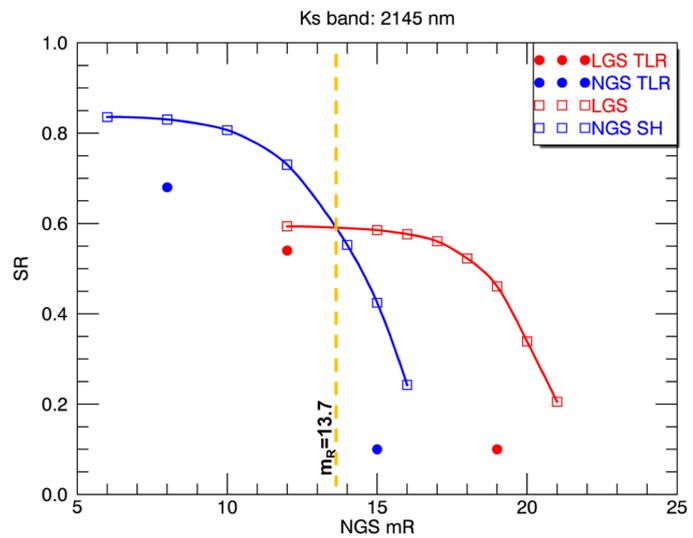

*Figure 7 Estimated performance in Ks-band of the ERIS AO system*

### 5.2 Control loops in NGS mode

Figure 6-right summarizes the system of control loops and LUTs in NGS mode. The scheme is similar to the LGS one disabling the loops that are LGS dependent.

## 6. AO SYSTEM PERFORMANCE

The ERIS AO system performance is computed in terms of Strehl ratio (SR) as a function of the NGS magnitude and it is shown in Figure 7 for the Ks band in comparison with the top-level requirements for both LGS and NGS modes. The curves have been obtained adding to the end-to-end numerical simulation results the wavefront error budget components that have not been directly included in the simulation. The numerical simulations have been provided by the PASSATA code developed at INAF-Arcetri[11].

The parameters of the numerical simulation are shown in Table 1 and the additional terms of the wavefront error (WFE) budget are the following:

- figuring error of DSM and M1 (after removing DSM correctable component)
- correction residual of Non Common Path Aberrations (NCPA)
- pupil registration loop residual
- interaction matrix calibration noise
- truth sensing residual error
- residual vibration compensation

*Table 1. Main simulation parameters*

| Parameter | Value |
|---|---|
| Seeing | 0.87 arcsec (at z=30°) |
| Outer scale | 22 m |
| Atmospheric layers | 10 (std. Paranal profile) |
| Wind speed | 12 m/s |
| NGS star type | G2 (spectra from Pickles) |
| Telescope entrance pupil | D = 8.12 m, 15.9% obs. H = 90 m below M1 |
| Laser | 1.2as WFHM including upward prop. Sodium profile "Single Peak" from Prommer&Hickson Simulated elongation and downward propagation |
| DSM | 1170 influence functions from FEA |
| WFS camera (CCD220) RON Excess noise Dark | EMCCD 240x240 pix 80e- rms/G (G=EM gain, 1≤G≤400) proper avalange statistics 3.6 e-/pixel/s (fs≥900) 1.8 e-/pixel/s (900>fs≥500) 1.4 e-/pixel/s (500>fs≥360) 1.08 e-/pixel/s (360>fs≥180) |
| LGS WFS subap FoV wavelength Pix/subap tranmission | SH 40x40 subap 5 arcsec $\lambda_{eff}$ = 589 nm (BW=40nm) 6x6 0.48 (including CCD QE) |
| HO/LO NGS WFS subap FoV wavelength Pix/subap tranmission | SH; HO:4x4 subap; LO:4x4 subap 2.5 arcsec $\lambda_{eff}$ = 768 nm (BW=600-1000nm) HO: 6x6; LO: 12x12 0.34 (including CCD QE) 107 detected ph/arcsec$^2$/m$^2$/s |
| RTC (SPARTA) Max frame frequency Centroid algorithm Control base Total effective delay (in frame units) | HO: 1kHz; LO: 500Hz Weighted CoG, gaussian weight map Karhunen–Loève modes fitted on the DSM influence functions 1 for $f$ < 334 Hz 2 for 334 Hz < $f$ < 667 Hz 3 for 667 Hz < $f$ < 1000 Hz |

The total contributions of the budget for the NGS and LGS modes are 110 nm and 150 nm rms WFE respectively. They are dominated by an extremely conservative contribution of the residual vibration compensation of 83 nm and 113 nm rms, based on vibration data in a different focal station (VLTI) and without considering the effect of the Active Vibration Cancellation algorithm developed and successfully tested by ESO.

For each NGS magnitude, both in NGS and LGS mode, the number of correcting modes and related loop gains, the loop frequency, the Gaussian size of the WCoG weighting maps and the CCD220 gain have been optimized.

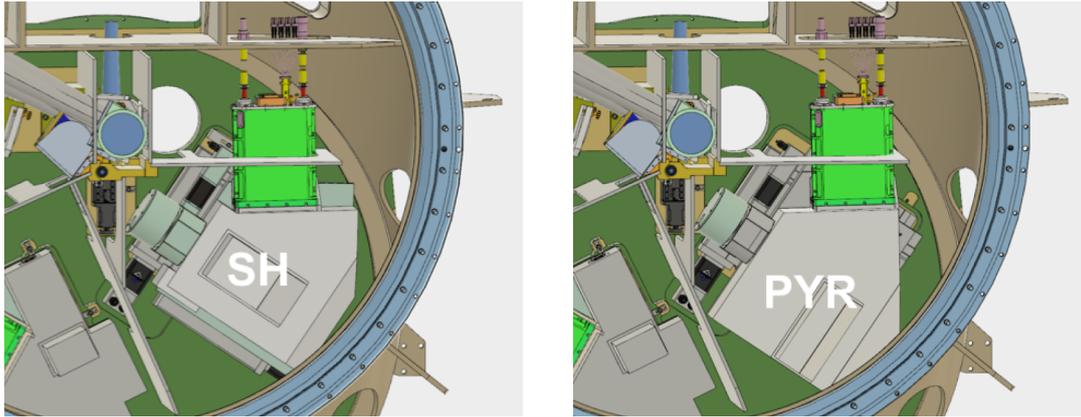

*Figure 8. Comparison between SH WFS (left) and Pyramid WFS (right).*

The performance shows a conservative SR(Ks)=83% in the on-axis bright end in NGS mode and 60% in LGS mode. For on-axis observations, the NGS magnitude for selecting the better performing mode (LGS/NGS) is $m_R$=13.7. The comparison with the TLR requirements show a contingency of 150nm and 100nm rms WFE in the bright end of the NGS ($m_R$=8) and LGS ($m_R$=12) mode, respectively.

## 7. THE PYRAMID WFS UPGRADE OPTION

Considering the performance success of the Pyramid WFS in other AO systems developed and experienced by the INAF-Arcetri AO group (e.g. FLAO at LBT[7] and MagAO at Magellan Telescope[8]), an upgrade option of the baseline SH WFS with a PWFS has been studied as part of the ERIS preliminary design. To be noted that both of the cited PWFS-based AO systems are coupled with a DSM sharing the same technology of the VLT-UT4.

### 7.1 Pyramid WFS upgrade opto-mechanics

The PWFS design option has the same opto-mechanical interfaces of the baseline SH WFS (see Figure 8). The internal components are shown in Figure 9, where the ones having labels with a thick frame share the same design with the SH WFS, at least for the mechanical part. As for the baseline SH sensor the patrolling of the FoV in transversal and longitudinal position is achieved by using a rotating periscope and an XY stage.

A different relay lens has been designed to accommodate a longer focal ratio at the pyramid (F/42.5 instead of F/20). The relay lens creates a pupil image of 16mm over a fast steering mirror that produces the fast angular modulation of the PSF.

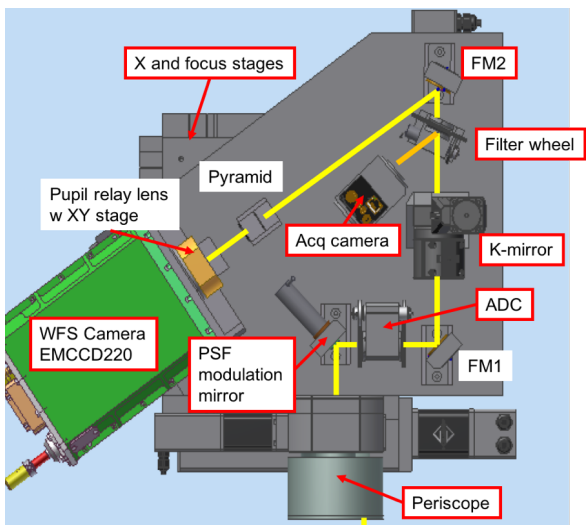

*Figure 9. Pyramid WFS components*

The angle of incidence on this mirror is 45 degree. Close to it, a double counter-rotating prism acts as ADC. This is based onto a similar design of the SH NGS WFS (see Section 4.8.3), but the larger pupil diameter requires smaller apex angles inside the prisms. The derotator prism is placed between the ADC and the focal plane.

After the derotator, a dichroic beam splitter reflects shorter wavelengths for target acquisition to an acquisition camera. Due to the very long focal ratio, a focal reducer doublet is placed after the dichroic to shorten the focal ratio to F/20 and match the same configuration of the SH NGS WFS.

On the other arm of the beam-splitter, the transmitted light is focused on a double-pyramid[12] placed together with a field stop at the F/42.5 focal plane. Finally, a pupil relay lens (a triplet) reimages the four pupils onto the detector. Pupil images have the correct sampling of 40x40 pixels. The compactness of the PWFS signal on the CCD area could allow to increase the number of sub-apertures up to the limit of the SPARTA RTC capabilities to even better reduce the aliasing error contribution in the reconstruction process.

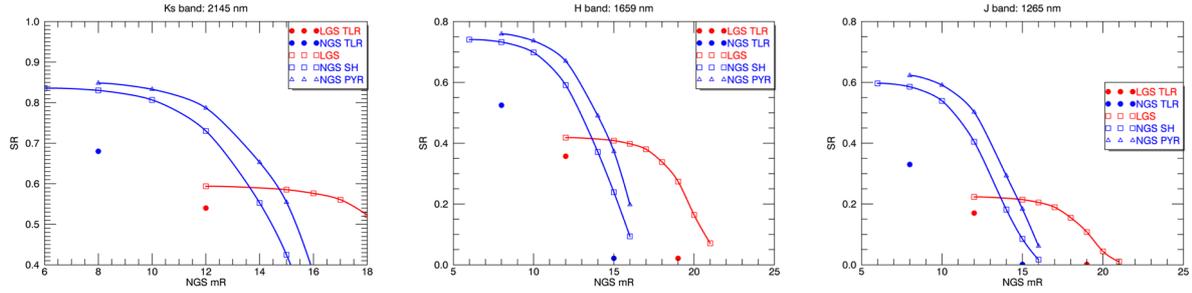

*Figure 10 Comparison between performance of SH and Pyramid WFS in the NGS-mode for Ks (left), H (center) and J (right) bands*

The NGS LO configuration can be achieved by switching to a different camera lens set-up reducing the pupil images diameter to avoid summing up dark current and RON contributions. Optical design for such configuration will be studied in the final design phase.

### 7.2 Pyramid WFS upgrade performance

Figure 10-left compares the numerical simulations of the PWFS with respect to the SH WFS in the NGS-mode. The simulation parameters are the same of the Table 1 with the additional parameter of the PSF modulation amplitude that ranges between 2 and 3 $\lambda$/D. The same conservative contribution of the WFE budget reported in Sec. 6 is applied to the PWFS, compressing the SR performance in the bright-end.

The PWFS increases the SR(Ks) at the crossing magnitude from 60% to 70% with a corresponding fainter crossing of the LGS curve of about 1 magnitude. Figure 10-center and right show the corresponding results in J and H band.

The increase of PWFS SR with respect to SH WFS produces a larger fraction of energy in the central peak that turns in a reduction of exposure time to obtain the same SNR in background limited observations. Figure 11-left reports the exposure time reduction factor, reaching a value more than 2 in J band.

The same increase of SR can be used with the PWFS to have the same SR results of the SH WFS at a given seeing for a larger value of seeing and, consequently, for a larger fraction of nights. Figure 11-right shows the seeing (solid line) for which the PWFS produces the same SR of the SH WFS when the latter is performing with a median seeing of 0.87as. The PWFS can produce the same results with seeing above 1as.

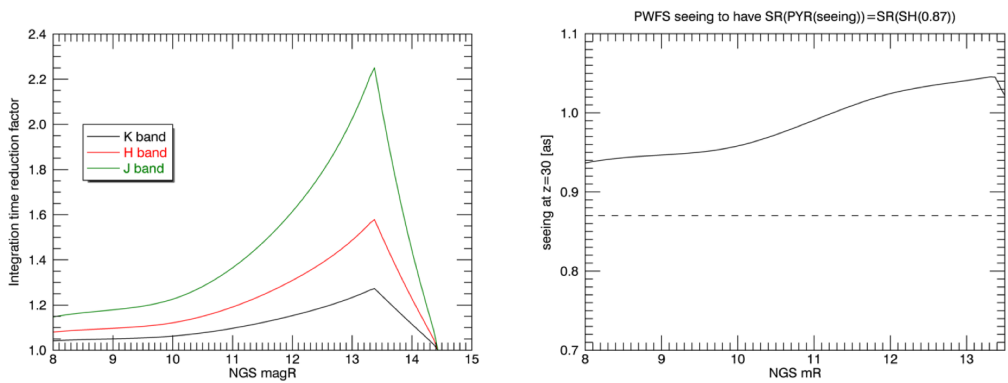

*Figure 11 Left: exposure time reduction factor for background limited observations of PWFS with respect to SH WFS. Right: seeing used by PWFS (solid curve) providing the same SR of the SH WFS case with 0.87as seeing (dashed curve).*

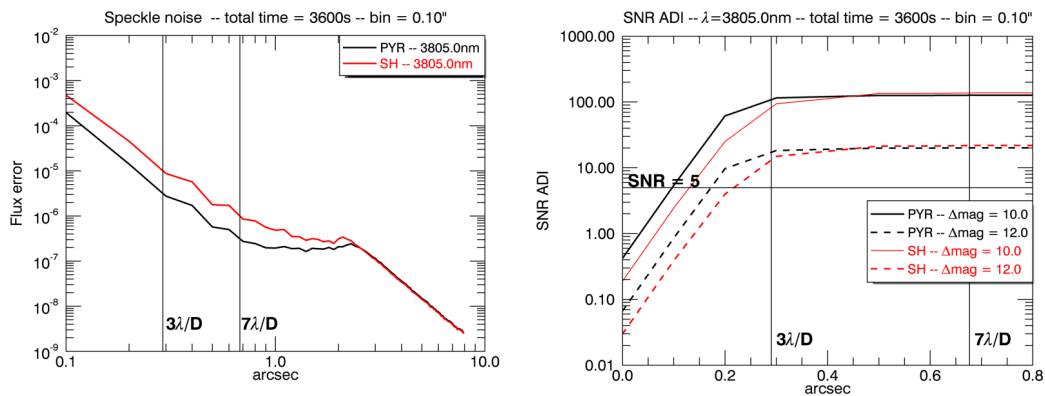

*Figure 12 Left: Radial profile of the speckle noise (see text). Right: detection SNR vs distance from central star (see text).*

The high contrast applications show even more interesting results, because of the well-known ability of the PWFS in a better rejection of the aliasing error. Figure 12-left shows the radial profile of the speckle noise computed from rms of difference of uncorrelated pairs of L'-band PSFs with 1.6s exposure time each, during AO bright-end on-axis correction. No coronagraphy is used in the simulation and the PSF binning has been set to $0.1 \times 0.1 as^2$. The PWFS speckle noise is about 3 times less than SH in the TLR requirement area between 3 to 7 $\lambda/D$. The curves, of course, superimpose beyond the AO control radius. The speckle noise is the dominant term below 3 $\lambda/D$ in the SNR of detection of a faint companion for a simulated ADI observation, while the background noise dominates for larger radial distance. We assumed a background brightness $m_{L'}$=3.9 mag/arcsec$^2$[13] in the simulation. Figure 12-right reports the detection SNR in 1h of a $\Delta m_{L'}$=10 (TLR requirement) and $\Delta m_{L'}$=12 companion for both Pyramid and SH WFS without coronagrafy. The SNR threshold required by TLR is SNR=5 in 1h showing that the PWFS pushes the performance in a closer region (around 2 $\lambda/D$) with fainter companion ($\Delta m_{L'}$=12) with respect to the minimal TLR requirements.

## 8. CONCLUSIONS

The ERIS Consortium has taken the lead of the 1-5μm ERIS instrument development beginning 2015. The Preliminary Design Review has been successfully met in February 2016 and now the project is in the Final Design Phase.

The ERIS AO system design makes a large re-use of AOF technology to save development and implementation time: the Deformable Secondary Mirror, the 4LGS Facility, the WFS Cameras (EMCCD) and the SPARTA RTC.

The pre-optics design, based on flat or almost-flat optics, has driven by providing high stability of NCPA and alignment decoupling with respect to the science instruments NIX and SPIFFIER.

The required observing modes (LGS, NGS and Seeing Enhanced) are implemented with the use of two SH WFSs units. The on-axis LGS-only WFS board implements the 40x40 SH with 5as FoV and is located on a linear stage for Na layer tracking. The NGS-only WFS is able to switch between a HO (40x40 SH, 2.5as FoV) and a LO (4x4 SH, 2.5as FoV) configuration for the NGS and the LGS mode respectively. The LO configuration provides also the truth sensing function for the LGS WFS focus and slope offset tuning. The NGS WFS board is located on a dual-axis stage to patrol the R=1arcmin acquisition field (together with the on-board periscope) and the compensation of the differential focus with respect to the science instruments.

The numerical simulations and analysis show that the SR requirements required by the TLRs are met with a contingency of 150nm and 100nm rms WFE in the bright end of the NGS ($m_R$=8) and LGS ($m_R$=12) mode, respectively.

In order to push the performance of the ERIS AO, especially in the high-contrast regime, an upgrade of the NGS WFS to the Pyramid WFS has been proposed. The PWFS unit has been designed to have the same opto-mechanical interfaces as the SH WFS with sensitive better SR performance of SH WFS, especially at short wavelengths (J and H bands). In the high contrast applications allows to push the faint companion detection performance for inner angles (2 $\lambda/D$) with fainter companions ($\Delta m_{L'}$=12).

# REFERENCES

[1] Madec, P.-Y., *et al.* "Adaptive Optics Facility: control strategy and first on-sky results of the acquisition sequence," Proc. SPIE 9909, in this proceedings (2016).
[2] Briguglio, R., Biasi, R., Xompero, M., Riccardi, A., Andrighettoni, M., Pescoller, D., Angerer, G., Gallieni, D., Vernet, E., Kolb, J., Arsenault, R., Madec, P.-Y., "The deformable secondary mirror of VLT: final electro-mechanical and optical acceptance test results," Proc. SPIE 9148, p. 914845 (2014).
[3] Hackenberg, W. K., *et al.* "ESO 4LGSF: Integration in the VLT, Commissioning and on-sky results," Proc SPIE 9909, in this proceedings (2016)
[4] Taylor, W. D., *et al.* "NIX, the imager for ERIS: the AO instrument for the VLT," Proc. SPIE 9909, in this proceedings (2016).
[5] George, E. M., *et al.* "Making SPIFFI SPIFFIER: upgrade of the SPIFFI instrument for use in ERIS and performance analysis from re-commissioning," Proc SPIE 9909, in this proceedings (2016).
[6] Dolci, M., *et al.* "Deign of the ERIS calibration unit," Proc. SPIE 9909, in this proceedings (2016)
[7] Esposito, S., Riccardi, A., Pinna, E., Puglisi, A., Quiros-Pacheco, F., Arcidiacono, C., Xompero, M, Briguglio, R., Agapito, G., Busoni, L, Fini, L., Argomedo, J, Gherardi, A, Brusa, G., Miller, D., Guerra, J. C., Stefanini, P., Salinari, P., "Large Binocular Telescope adaptive optics system: new achievements and perspectives in adaptive optics," Proc. SPIE 8149, p. 814902 (2011).
[8] Morzinski, K., *et al.* "MagAO: status and science," Proc. SPIE 9909, in this proceedings (2016)
[9] Pfrommer, T., Hickson, P., "High resolution mesospheric sodium properties for adaptive optics applications," A&A 565, p. A102 (2014)
[10] Muradore, R., Pettazzi, L., Fedrigo, E., Clare, R., "On the rejection of vibrations in Adaptive Optics Systems," Proc. SPIE 8447, p. 844712 (2012)
[11] Agapito, G., Puglisi, A. and Esposito, S., "PASSATA: object oriented numerical simulation software for adaptive optics," Proc. SPIE 9909, in these proceedings (2016).
[12] Tozzi, A., Stefanini, P., Pinna, E., and Esposito, S., "The double pyramid wavefront sensor for LBT," Proc. SPIE 7015, 701558, (2008).
[13] https://www.eso.org/gen-fac/pubs/astclim/paranal/skybackground/